\documentclass[12pt,aps,prb,preprint]{revtex4}

\usepackage{amsmath}
\usepackage{amsfonts}
\usepackage{amssymb}
\usepackage{graphicx}
\usepackage{epsfig}

\begin{document}

\bibliographystyle{unsrt}

\title{Influence of a humidor on the aerodynamics of baseballs}
\author{Edmund R. Meyer}\email{meyere@murphy.colorado.edu}
\affiliation{JILA, NIST, and Physics Dept., Univ. of CO, Boulder, CO
80309}
\author{John L. Bohn}
\affiliation{JILA, NIST, and Physics Dept., Univ. of CO, Boulder, CO
80309} 
\date{\today}

\begin{abstract}
We investigate whether storing baseballs in a controlled humidity
environment significantly affects their aerodynamic properties.  To do
this, we measure the change in diameter and mass of baseballs
as a function of relative humidity (RH) in which the balls are stored. 
We then model trajectories 
for pitched and batted baseballs to assess the difference between those 
stored at 30\% RH versus 50\% RH.  The results
show that a drier baseball may be expected to curve slightly more
than a humidified one for a given pitch velocity. We also find that the 
aerodynamics alone would add ~2 feet to the distance a moister ball is hit. 
However, this is compensated by a ~6 foot reduction in batted distance due to 
the well known change in coefficient of restitution of the ball. 
We discuss consequences of these results for baseball played at Coors Field 
in Denver, where baseballs have been stored in a humidor at 50\% RH since 2002.
\end{abstract}

\maketitle

\section{Introduction}

The game of baseball is strongly influenced by the aerodynamics of
its central object, the ball.  The very fact that the ball travels
through the atmosphere is enough to make a typical home run ball hit at 
sea-level travel $400$ feet rather than the $\sim 750$ feet it would 
travel in vacuum\cite{adair}. Vice versa, that same $400$ foot home run 
might be expected to go $420$ feet in the thin air of Denver, one mile above 
sea-level. As noted by Chambers and co-workers\cite{chambers}, this is 
less than one might expect from aerodynamic considerations alone due to the  
prevailing northeast winds in Denver. 
Because the ball is expected to travel farther, the idea of introducing a major
league team (The Colorado Rockies) to the Denver area was met with opposition
in the early 1990's. Indeed, hitters counted on the so-called ``Coors
Field advantage'' to help boost their hitting numbers, such as home
runs (HR), runs batted in (RBI) and batting average (BA). On the other 
hand, pitchers dreaded throwing in Denver because the thin air was
blamed for the lack of break of a curveball or slider, which are common 
tricks of the trade for any modern pitcher.

In 2002 the Rockies organization attempted to mitigate these effects
by storing the balls in a humidity-controlled environment.  The
conventional wisdom held that a light and dry ball travels farther
than a heavy and moist ball.  Coors Field engineer Tony Cowell
reported that baseballs stored in the dry atmosphere of Denver were
both lighter and smaller than the weight and circumference specified
by the rules of Major League Baseball\cite{dan}.  So, starting in
the 2002 season, the Rockies have stored their baseballs in a
humidor behind the beer storage facility.  This sealed room is
maintained at a temperature of 70$^\circ$ F and 50\% relative
humidity (RH), consistent with the specifications of baseball
manufacturer Rawlings.

The use of the humidor remains a matter of controversy within
professional baseball, and a matter of fascination for the public
that follows the sport.  It is widely believed that Coors Field has
yielded fewer home runs and fewer runs altogether in the humidor era.
Indeed, the statistics for various offensive and defensive
benchmarks at Coors Field do support a small but measurable
correlation with the presence of the humidor, as shown in Table
\ref{data}. This table shows the various statistics for games played
at Coors Field as compiled in the seven seasons before the humidor
was installed (BH), and compared to the same statistics from the first
five seasons after (AH). For comparison, the same numbers for the
entire National League (NL) are presented in the same two time
frames. These data were obtained from Refs.\cite{rmn,bbr} except average fly-ball 
distances. These were obtained from Ref.\cite{mgl}, which only reports the two 
years BH and 5 AH.

\begin{table}[ht]
  \begin{center}
    \begin{tabular}{lc|c|c}
      \hline
      & BH & AH & BH $-$ AH\\
      \hline
      Rockies ERA @ Coors & 6.14(52) & 5.34(52) & .80(74)\\
      \hline
      NL ERA @ Coors & 6.50(47) & 5.46(48) & 1.04(67)\\
      \hline
      NL Avg. ERA & 4.37(18) & 4.28(12) & 0.09(22)\\
      \hline
      HR$/$Team-Game @ Coors& 1.59(19) & 1.26(18) & 0.33(26)\\
      \hline
      NL Avg. HR$/$Team-Game & 1.06(09) & 1.05(04) & 0.01(10)\\
      \hline
      Runs$/$Team-Game @ Coors & 6.94(49) & 5.87(36) & 1.07(61)\\
      \hline
      NL Avg. Runs$/$Team-Game & 4.7(19) & 4.58(12) & 0.18(22)\\
      \hline
      Avg. Fly Ball Distance & 323(4) & 318(3) & 5(3.5)
    \end{tabular}
    \caption{\label{data} Examples of run production numbers at Coors Field
      before (BH) and after (AH) the inception of the humidor in 2002. These 
      are compared to the NL averages in the same two periods. Compare the NL 
      averages to the Coors averages and note the deviation from zero.  The 
      quoted uncertainties are calculated using the year-to-year variance, as 
      the counting statistics contribute negligibly to this uncertainty.}
  \end{center}
\end{table}

In all cases, there has not been a significant change for the NL as
a whole.  This fact presumably represents an averaging over a host
of other changes that may have taken place from one half-decade to
the next. In all the categories for statistics at Coors Field,
however, there is a definite trend toward lower ERA's and fewer
runs (indicating more successful pitching), 
by more than one standard deviation.  In a game as complex as
baseball, it is impossible to attribute a cause-and-effect
relationship between these results and the humidor.  For example,
Rockies pitchers may have been more talented since 2002; they may also have benefited 
from a strike zone enlargement in 2001\cite{mlb}; however, consequences 
of these changes are difficult to quantify by means other than the very
performance statistics we use to check the humidor's effect.

Nevertheless, one may wonder what possible quantifiable effects the
humidor might have on the game. As early as the late 1980's, the
president of the National League (NL), A. Bartlett Giamatti, asked
Robert Adair, a Yale physicist, to test the effect of a humidor on
baseballs. Adair found that under the most extreme conditions the
humidified balls gained weight and were less elastic\cite{adair}. 
Subsequent work by Kagan\cite{kag2} carefully measured the dependence of the ball's 
coefficient of restitution on RH. He then translated this into batted ball distance, 
finding that storing balls in an increased RH environment reduces the distance a
well-hit ball travels by $3$ feet for every 10$\%$ change in RH.\cite{kag2} This would 
account for a 6 foot reduction in fly ball distances at Coors Field after the humidor 
was installed. Another possible effect of the humidor is to make the hide
surface of the balls more supple, enabling the pitchers to maintain
a better grip and therefore pitch as they would with balls stored in higher
RH elsewhere. This effect has been reported by pitchers, but
remains unexplored quantitatively.

Kagan's work\cite{kag1,kag2} focused on the influence of RH on the bat-ball 
collision. In this article we expand this work to investigate how storing a ball
in a humidor affects the aerodynamics of the ball in
flight. To do so, we perform measurements of the change in diameter
and mass of baseballs as a function of RH. Strikingly, we
find that this variation is less than the variation already allowed
by the rules of Major League Baseball (MLB).  Nevertheless, the lightest,
smallest allowed balls, if dried out, will exceed the limits in the
rulebook.  To assess the effect of these changes in the ball's
flight, we numerically solve the equations of motion for pitched and
batted balls, including lift and drag forces, and the dependence of
these forces on the speed and spin rate of the ball.  We find that,
on average, curveballs break slightly {\it more} for dried baseballs
than for humidified baseballs, and that batted balls travel slightly
{\it less far} when they are dry than when they are humidified for given 
initial trajectories. Therefore, the combined effect of RH on the elasticity and 
aerodynamics of the ball suggest that post-humidor batted balls travel perhaps 
4 feet less than dry baseballs.

The rest of the paper is outlined as follows. Sec. \ref{exp} will
provide the results of an experiment designed to measure the effect of RH on
the size and mass of the ball. Sec. \ref{aerodyn} discusses the 
approximate models of aerodynamic forces acting on a baseball in flight. 
Sec. \ref{numerics} evaluates the effects of RH on the 
aerodynamics for curve balls and well-hit baseballs. In Sec. \ref{con} 
we summarize our results and present possibilities for future work.

\section{Experiment}\label{exp}

To assess the effect of RH on the baseballs, we measured the
mass and diameter of a collection of balls stored at various
RHs.  We stored five Major League baseballs each in airtight
containers in which the RH was held at the
constant values 32$\%$, 56$\%$, and 74$\%$.  These humidities were
maintained by including saturated salt-and-water solutions inside
the containers.\cite{NISTpaper} The RH in each box was monitored and
found to hold constant to within $\pm 1 \%$, which was also our
approximate measurement uncertainty.  The temperature was not
carefully controlled, but remained near 70$^{\circ}$ F during 
throughout the experiment.

The balls' masses were measured to $\pm 0.1$g using a digital
balance.  Ball diameters were measured by placing them on a flat
marble slab and measuring the top of the ball using an accurate
height gauge.  Balls were marked so that the same diameter could be
reliably measured more than once.  Measurement uncertainties of the
diameters were found to be $\pm 0.013$ in.

The diameter of each ball was measured across five different orientations. 
Three of these were on approximately mutually orthogonal directions on the
leather surface. The other two diameters were across the seams,
hence slightly larger than the leather diameters. Variations 
of the measured diameters from one ball to the next, and even from one axis
to another on the same ball, were larger than the measurement
uncertainty. Because of these variations, direct comparisons between the
diameters of dry and wet baseballs do not show the effect of RH. 
We therefore report the data as the {\it fractional
change} in diameter from that of a dry ball held at 32 $\%$ RH,
$d/d_{\rm dry}$.  Based on this measurement, we are unable to
distinguish a difference in the expansion of the ball's diameter
measured on the leather as opposed to on the seams.  We therefore
averaged all five measurements of $d/d_{\rm dry}$ for each ball, and all 25
measurements for the five balls in each container.

The experiment began with all balls held in a dry environment for a
period exceeding two months.  Two sets of balls were then moved into
the ``humid'' (56$\%$ RH) and ``wet'' (74$\%$ RH) containers.  The
quantities $d/d_{\rm dry}$ and the mass ratio $m/m_{\rm dry}$ were
recorded at weekly intervals, showing that they saturated on a
timescale of $\sim 2$ weeks. For comparison, the balls at Coors Field are 
assumed to be stored for timescales long compared to this. The humidor 
stores 400 dozen baseballs and the stock is rotated bringing the 
oldest balls out first, as reported in Ref.\cite{rmn}.

\begin{figure}[h]
  \begin{center}
    \includegraphics{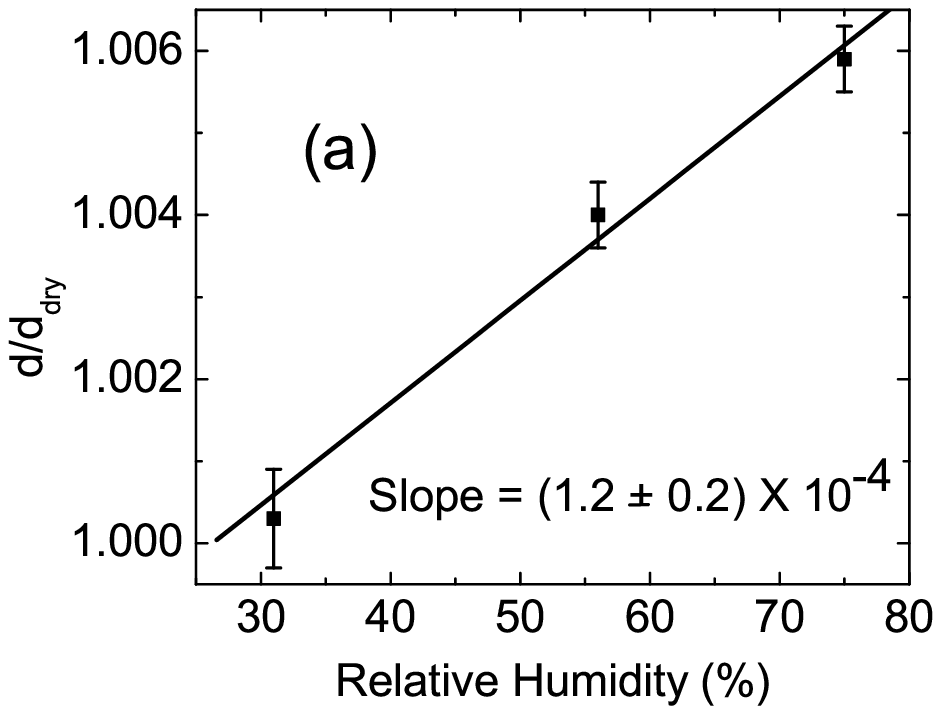}
    \includegraphics{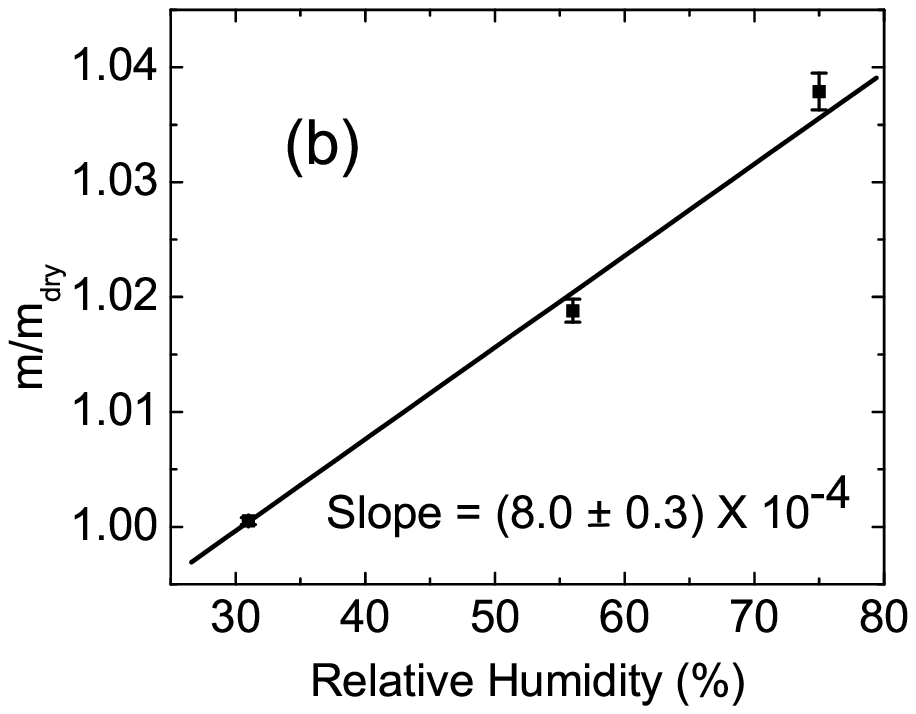}
    \caption{\label{exp_results} (a) Relative diameter and (b) 
    relative mass as a function of relative humidity (RH).}
  \end{center}
\end{figure}

The steady-state dependence of ball size and mass with humidity is
reported in Fig. \ref{exp_results}.  A linear fit to the data show that a given
diameter of a given ball can be expected to increase by $0.012\%$ 
for each percent of RH, or by about $0.24 \%$ upon changing
the humidity from 30$\%$ (typical of summer weather in Denver) to
$50\%$ (the specification of the humidor).  Likewise, the mass of
the balls increases by $0.08\%$ for each percent RH, or
$1.6 \%$ between 30$\%$ RH and 50$\%$ RH.  A consequence of this is
that humidifying the balls increases the density of the balls by
$0.9 \%$.

\begin{figure}[h]
  \begin{center}
    \includegraphics{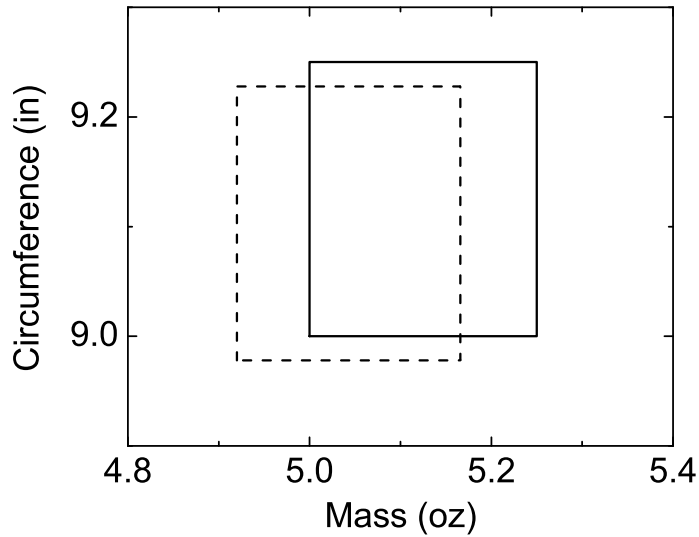}
    \caption{\label{var_all} The solid line represents allowed variations of ML 
      baseballs, while the dashed line represents these same dimensions 
      shifted to account for a 20\% reduction in RH.}
  \end{center}
\end{figure}

Fig. \ref{var_all} summarizes the difference the humidor makes in the size
of the ball.  On one axis of this plot is the circumference of the
balls, on the other is the mass.  The region indicated by the solid
lines shows the allowed limits on circumference and mass specified
by the rules of MLB.  By contrast, the dashed line
shows what these specifications become if the ball is dried to
30$\%$ RH from $50\%$ RH.  Strikingly, drying out a baseball
produces a smaller change than, say, substituting the smallest,
lightest ball for the largest, heaviest one allowed by the rules.

Armed with these data, the calculations presented below assess the
difference between a ``dry'' ball and a ``humid'' one, i.e., the dry
ball is by definition $0.24 \%$ smaller in diameter and $1.6 \%$
lighter than the humid one.

\section{Aerodynamic Forces}\label{aerodyn}

\begin{figure}[h]
  \begin{center}
    \includegraphics{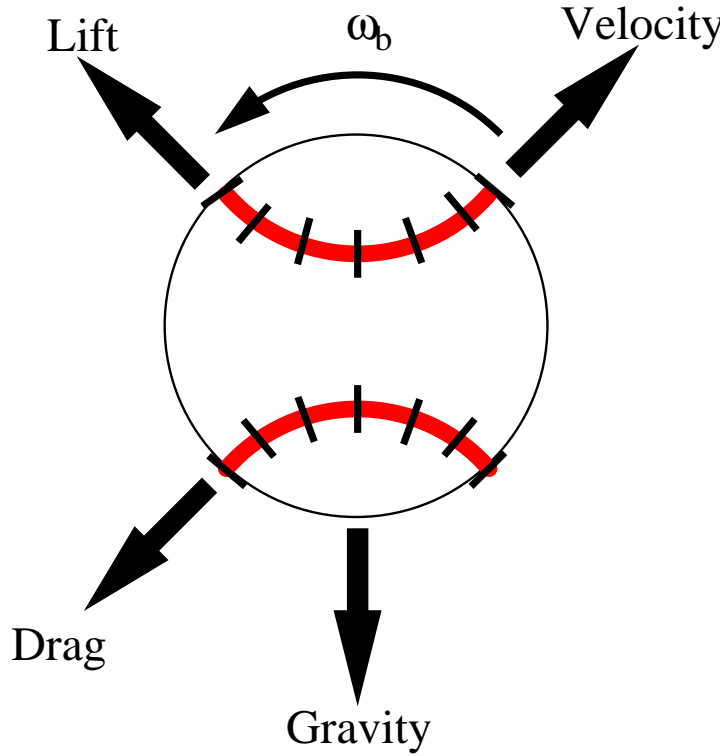}
    \caption{\label{forces} Aerodynamic and gravitational forces acting on 
      a baseball.}
  \end{center}
\end{figure}

The aerodynamics of balls in flight have been studied since the time
of Newton, who was the first to appreciate the Magnus force
responsible for the curve of a tennis ball\cite{newton}.  
This force, along with the drag force, determine the
trajectory of the ball in flight.  A free-body diagram of a baseball
in flight is shown in Figure \ref{forces}. As shown, a baseball
moving to the right with a counterclockwise spin (i.e., a backspin)
experiences a lift force with an upwards vertical component.

While these forces are understood qualitatively for baseballs, their
details remain somewhat obscure. In this section we summarize
existing measurements of drag and lift forces for baseballs, and
construct models that we will use in our trajectory simulations,
below.

\subsection{Drag}

The drag force originates from the air displaced by the ball as it
travels. It is therefore proportional to the cross sectional surface
area of the ball, $A$, and to the density of the air it displaces,
$\rho$. From dimensional considerations, the drag force must then
depend on the square of the velocity. Thus,
\begin{equation}\label{drag}
  {\bf D} = -\frac{1}{2}\rho C_D A v^2 \hat{\bf v},
\end{equation}
where the dimensionless drag coefficient $C_D$ characterizes the
strength of the drag force under particular circumstances. The direction 
of the drag force is such that it opposes the motion of the ball.

This form of the drag force assumes an aerodynamic regime where the 
fluid's viscosity is low, or equivalently where the Reynolds number is
large.  The Reynolds number is defined by
\begin{equation}
  \mathcal{R} = \frac{v d}{\nu},
\end{equation}
where $v$ is the ball velocity, $d$ the diameter of the ball, and
$\nu$ the kinematic viscosity of the fluid.  Baseball is played in a
regime where $\mathcal{R} \sim 10^4-10^5$, and viscosity plays a
fairly small role. For lower velocities, energy is dissipated into
the environment, leading to a drag coefficient that diverges as
$1/v$.\cite{landl} For baseballs in flight, $C_D$ as defined is
roughly constant with velocity.

However, the drag coefficient for a spherical object changes
dramatically in a certain range of $\mathcal{R}$.  Near this
``drag crisis,''  $C_D$ can drop by a factor of 2--5 over a narrow
range of $\mathcal{R}$.\cite{landl} This reduction is due to a
change from laminar flow (i.e. smooth flow) to turbulent flow. The
wake region formed behind the ball begins to shrink due to boundary layer 
separation. Eddies build up in the wake region causing pressure to increase on
the side of the ball opposite the direction of travel. The reduction
in $C_D$ is so great that it actually diminishes the total drag
force, despite the quadratic dependence on velocity, for
$\mathcal{R}$ on the order of $10^5$.

Frohlich\cite{froh} noted that the game of baseball is played in
the regime where the drag crisis plays a crucial role. In addition,
using the results of Achenbach\cite{achen} on sand roughened
spheres, Frohlich pointed out that the roughness of a baseball
affects the Reynolds number at which the crisis occurs. Namely, a
rougher ball reaches the onset of the drag crisis for a smaller
value of $\mathcal{R}$, hence a lower velocity. For a baseball,
Achenbach defined the roughness using the average height of the seam 
above the hide relative to the diameter of the baseball hide. 
This quantity is defined as $\epsilon$. As the ball rotates,
this disparity can appear on average as a ``sand-roughened'' sphere.
Frohlich also pointed out that if a hitter can ``punch'' through the
crisis, i.e., get the ball to exceed the speed where the crisis
occurs, then the ball would travel a greater distance when struck
than expected using a constant drag profile.

\begin{figure}[h]
  \begin{center}
    \includegraphics{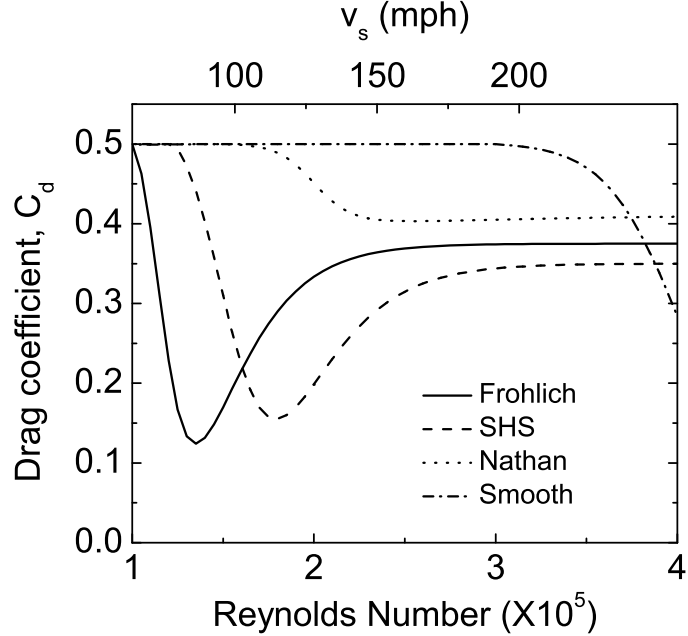}
    \caption{\label{crisis} Drag coefficient $C_D$ as a function of $\mathcal{R}$.
      On the top axis $\mathcal{R}$ is translated into a velocity for a standard 
      ball (5.125 ounces, 9.125 inches in circumference) at representative Denver air 
      pressure and kinematic viscosity. Each curve represents an approximate fit to 
      the data from a given publication.}
  \end{center}
\end{figure}

Sawicki, Hubbard and Stronge (SHS)\cite{shs} extracted information
on $C_D$ versus $\mathcal{R}$ from data taken at the 1996 Atlanta
Olympic games\cite{alaw}. They found a possible occurrence of the
drag crisis. By contrast, experimental results of Nathan, {\it et
al.}\cite{nat2} suggest that the drag crisis is less dramatic for
baseballs than the SHS analysis.  Because of this discrepancy, we
perform trajectory simulations using several models of drag.  Fig.
\ref{crisis} presents the four drag coefficient profiles considered
in this paper, including a smooth ball that is not expected to
represent a real baseball.  The curve attributed to Frohlich\cite{froh}
represents a surface roughness of about $\epsilon=5\times10^{-3}$.

As a point of reference, in Fig. \ref{crisis} $\mathcal{R}$
has been converted into $v_s$, which is the velocity of a standard
baseball (5.125 ounces, 9.125 inches in circumference) in the atmosphere 
of Denver. Denver is located one mile above sea-level 
and therefore has a lower air density and a higher kinematic viscosity. 
A drag crisis that would occur at $70$--$100$ mph at sea-level would instead 
occur at speeds near $100$--$130$ mph in Denver. Thus, the drag crisis 
is likely a bigger factor for baseball played at sea-level than in Denver. 
We have shifted Nathan's published data\cite{nat2} to account for air density 
at Denver. We stress that these curves are approximate fits to published
data sets and shifted according to the value of $\nu$ and $d$ in the
$\mathcal{R}$-value. Nevertheless, the general features and differences 
between the curves are represented. For analytical convenience,  we have 
represented the drag coefficients in Fig. \ref{crisis} using the following 
functional form:
\begin{equation}
 C_D = a + b \tanh\left(\frac{\mathcal{R}-\mathcal{R}_d}{\Delta_d}\right) +
 c \tanh\left(\frac{\mathcal{R}_u-\mathcal{R}}{\Delta_u}\right),
\end{equation}
where $\mathcal{R}_d$ ($\mathcal{R}_u$) is the location of the drop
(rise) in the drag coefficient and $\Delta_d$ ($\Delta_u$) is the
corresponding width.  The data are consistent with two constraints:
for $\mathcal{R}\ll 10^5$, $C_D\approx 0.5$ and for $\mathcal{R}\gg
10^6$, $C_D\approx$ constant. Therefore, the fit has 7 parameters subject 
to two constraints given by
  \begin{equation}
    a - b \tanh\left(\frac{R_d}{\Delta_d}\right)+
    c\tanh\left(\frac{R_r}{\Delta_r}\right) = 0.5
  \end{equation}
  \begin{equation}
    a + b - c = \text{\rm Constant}.
  \end{equation}

\subsection{Lift}

The lift force plays a comparatively minor role in determining the
distance of a batted baseball,  but it is very important for pitched
baseballs. Lift is responsible for the break of the curveball and the
slide of the slider. Like drag, the lift force is
also proportional to the cross-sectional area and the density of
air. However, the direction of the force is not in the direction of
the drag force, but rather perpendicular to the direction of motion.
This is because the rotation of the ball serves to shift the wake region from
directly behind the ball, resulting in a force directed oppositely
to the wake region. The lift force is given by
\begin{equation}\label{lift}
  {\bf L} = -\frac{1}{2}\rho C_L A v^2 \;\hat{\bf v}\times\hat{\bf \omega}_b.
\end{equation}

The lift coefficient $C_L$ does not depend strongly on the value of 
$\mathcal{R}$ since it arises from rotation shifting the wake
region, not the size of the wake region or the eddies forming within
it. Adair\cite{adair} uses a differential model of $C_D$ to
determine an approximate $C_L$. However, careful measurements by SHS\cite{shs} and
Nathan\cite{nat2} find that $C_L$ depends linearly on the spin
parameter $S$, which is given by
\begin{equation}\label{sparam}
  S = \frac{r\omega_{\rm b}}{v},
\end{equation}
where $r$ and $v$ are the radius and velocity of the ball
respectively and $\omega_{\rm b}$ is the rotation rate of the ball
in radians per second. Baseball is played in a regime where $S$ lies
in the range $0<S<1/2$, where the lift coefficient's dependence on
$S$ is approximately\cite{shs}
\begin{equation}\label{cl}
  C_L = \left\{\begin{array}{cc}1.5 S, & S<0.1\\
      0.09 + 0.6S, & S>0.1\end{array}\right.,
\end{equation}
In addition, the orientation of the seams has a noticeable effect,
according to the unpublished work of Sikorsky and Lightfoot
referenced in Alaways.\cite{alaw}  We ignore this effect in the
present work. 

\section{Trajectory Calculations}\label{numerics}

The equation of motion for the baseball in flight is given by
\begin{equation}\label{eom}
  \dot{\bf v} = -{\bf g} + \frac{1}{m}\left({\bf D} + {\bf
  L}\right),
\end{equation}
where ${\bf v}$ is the ball's velocity, $m$ its mass, $\rm{\bf g}$ is
acceleration due to gravity, and the aerodynamic forces ${\bf D}$
and ${\bf L}$ are given in Eqs. \ref{drag} and \ref{lift}. For
simplicity, we restrict the ball's motion to a vertical plane, and
require the rotation axis to be orthogonal to this plane. The
trajectory then lies entirely in this plane.  All calculations assume a
``standard'' Denver atmosphere, with air density 0.91809 kg/m$^{\rm
3}$ (as compared to 1.0793 kg/m$^{\rm 3}$ at sea level) for a
temperature of 70$^\circ$ F.   We assume a kinematic viscosity for
Denver at this temperature of 2.095$\times 10^{\rm -5}$ m$^2$/s, as
compared to 1.8263$\times 10^{-5}$ m$^{\rm 2}$/s at sea-level.\cite{atmos}

Using this equation of motion, we assess the difference between dry
and humid baseballs with regard to two observable quantities.  The
first is the break of a pitched ball, i.e., the additional drop in
elevation of a ball with forward motion due to a downward lift force. 
The second is the distance a batted ball travels before hitting the 
ground. In both cases we explore the difference as a function of the 
ball's initial velocity.

\subsection{Aerodynamics of Pitched Baseballs}\label{pitches}

A typical MLB pitcher releases the ball from a height of $6.25$ feet
and a distance from the plate of $53.5$ feet at Coors
Field\cite{bbanal}. A claim that has been made is that curveballs
break more now that the balls have been stored in a humidity
controlled environment, bringing them closer to curveballs
in other Major League venues. We evaluate this claim by comparing
the relative arrival heights of curveballs launched horizontally with a
given velocity between $72$ mph (often called a ``slurve'') and $88$
mph (often called a ``power'' curve).

We consider two balls thrown with identical initial conditions.  The
first is a standard baseball with circumference 9.125 in and
mass 5.125 oz, i.e., the mean ball specified by the rules of
baseball. The second is this same ball, but stored in a dry
environment, reducing its circumference and mass by $0.24 \%$ and $1.6 \%$.  
We compute the difference in the heights of these balls upon reaching 
the plate, defined as $\Delta y = y^{\rm s} - y^{\rm d}$, where 
$y^{\rm s}$ is the height upon arrival of the standard ball, and 
$y^{\rm d}$ is the height of the dried ball.

\begin{figure}[h]
  \begin{center}
    \includegraphics{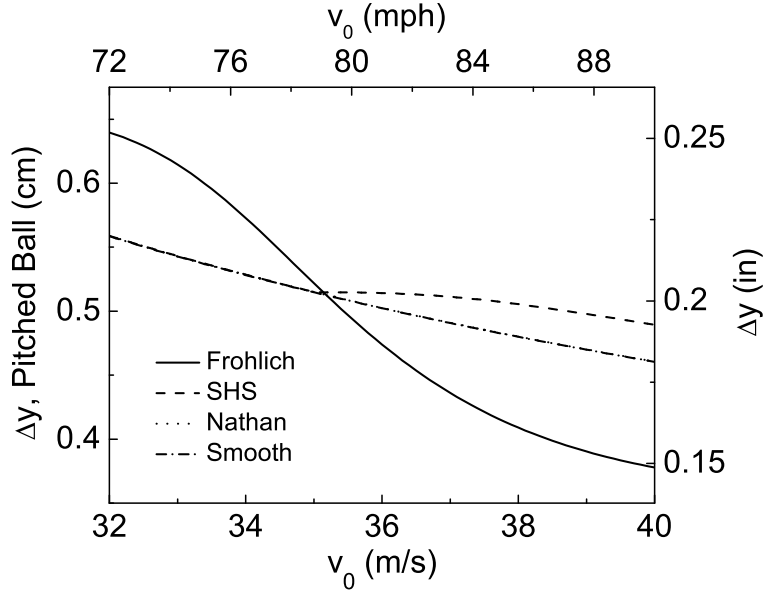}
    \caption{\label{breakvar} Relative break of a curveball thrown at Coors
      Field versus velocity. Positive values of $\Delta y$ mean that the ``drier'' baseball
      breaks more than the standard Rawlings baseball for a given initial velocity and 
      spin-rate. For ease of comparing to
      baseball units, the top axis is labeled in mph and the right axis in inches.}
  \end{center}
\end{figure}

This result is shown in Fig. \ref{breakvar}. For all initial
velocities, $\Delta y$ is positive, indicating that the
drier ball actually breaks more than the standard baseball. It is
moreover a small effect, changing the break of the ball by at most
0.25 in. This general result holds regardless of the specific drag
curve used. As the speed of the pitch is increased, the difference
in final height is diminished, and both balls break more similarly.
This is due primarily to the amount of time the ball is in flight: a
faster curveball will reach the plate sooner, thereby experiencing
the acceleration due to lift for shorter duration.

To see why the dry ball breaks more, it is important to understand 
the consequences of the equation of motion, Eq.\eqref{eom}. Defining 
the difference in accelerations due to lift, 
$\Delta a_{\rm L} = a_{\rm L}^{\rm s}-a_{\rm L}^{\rm d}$,
where s and d refer to standard and dried out baseballs, reveals the relative 
accelerations of the standard and dry baseballs. From 
Eq. \eqref{lift} the value of the acceleration due to lift is given by 
\begin{equation}\label{alift}
  a_{\rm L} = -\frac{1}{2}\frac{\rho A}{m}C_{\rm L} v^2,
\end{equation}
where the minus sign implies a downward direction. This is the total 
acceleration at the point of leaving the hand due to lift. If the value of 
$\Delta a_{\rm L}$ is positive the dry baseball experiences more acceleration due to lift. 
Since the initial pitch conditions are the same, the only parameters 
that change when the ball is dried out are $A$, $m$ and $C_{\rm L}$. The fractional 
change in $a_{\rm L}$, as the ball is thrown, is
\begin{equation}
  \frac{\Delta a_{\rm L}}{a_{\rm L}^{\rm s}} = \frac{\Delta C_{\rm L}}{C_{\rm L}^{\rm s}}
  + \frac{\Delta A}{A^{\rm s}} - \frac{\Delta m}{m^{\rm s}}.
\end{equation}
Since the direction is downward, if the value 
of $\Delta a_{\rm L}/a_{\rm L}^{\rm s}<0$, there is more break. Because $C_{\rm L}\sim d/2$ 
(see Eqs. (\ref{sparam}-\ref{cl})) and $A\sim d^2$, we can simplify the above 
expression to
\begin{equation}\label{fracalift}
  \frac{\Delta a_{\rm L}}{a_{\rm L}^{\rm s}} = 3\frac{\Delta d}{d^{\rm s}} - 
  \frac{\Delta m}{m^{\rm s}}.
\end{equation}
Based on the experimentally determined changes in diameter ($\Delta d/d^{\rm s}=0.24$\%) 
and mass ($\Delta m/m^{\rm s}=1.6$\%), we find that 
$\Delta a_{\rm L}/a_{\rm L}^{\rm s}\approx -0.88$\%. Therefore, the drier 
baseball experiences more lift acceleration and breaks more. While it is true that 
the standard ball experiences a greater lift force, this force is 
overcome by the ball's greater inertia due to its increased mass.

Nevertheless, reports from pitchers and batters alike assert that
the humidified balls break more. Our result shows that this cannot
be due purely to the aerodynamics. However, pitchers also report
that the humidified balls are easier to grip. It is possible that
they can put a greater spin on the ball. This would require Eq. 
\eqref{fracalift} to be modified to include a term $+\Delta\omega/\omega^{\rm s}$. 
Since the lift force is also proportional to spin, it would take only an 
additional spin of $\Delta\omega/\omega^{\rm s}=0.9 \%$ to overcome the 
aerodynamic effect arising from increased density of the humidified ball. 
Any additional spin the pitcher can provide beyond this value 
will further overcome the aerodynamic tendency of the standard ball to break 
less. It may be this effect on grip that is responsible for the ability of 
pitchers to perform at Coors Field more like they do elsewhere since the 
introduction of the humidor.

Another important item to note is the dependence on the type of drag
curve describing the crisis. Each curve is fairly consistent for
mid-range curveball speeds but it is clear that the sand-roughened
sphere of Frohlich shows a wider variation with speed. Also, for
higher speeds, the SHS drag curve would predict a larger break
variation than a smooth ball or one using the results of Nathan.
This points to the need to measure more accurately the drag curve, 
including its dependence on the spin of the ball.  Games played at 
sea-level are more susceptible to the subtleties
of the drag crisis due to the smaller value of $\nu$. A better 
understanding of the drag curve is imperative to understand quantitatively 
the slight variations of the game due to RH changes, or, indeed, due to the 
allowed variation in baseball dimensions.

\subsection{Aerodynamics of Batted Baseballs}

\begin{figure}[h]
  \begin{center}
    \includegraphics{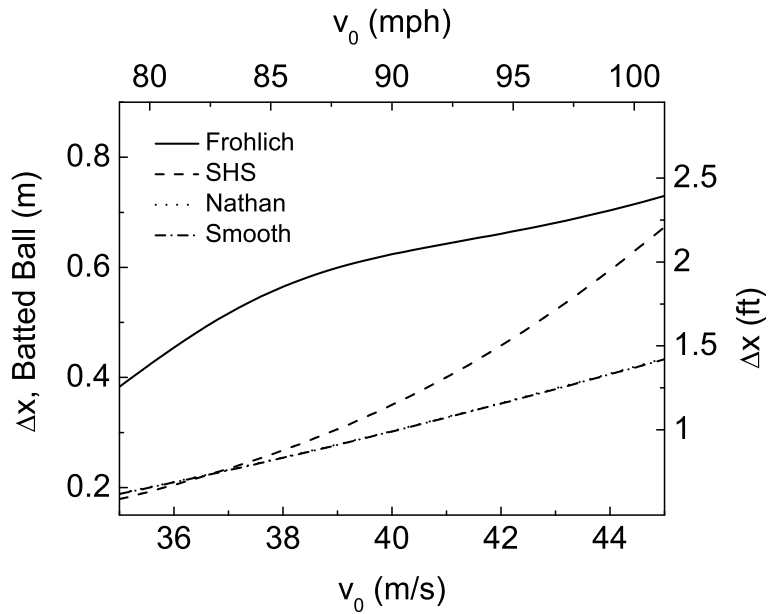}
    \caption{\label{rangevar} Relative variation in the range of a well-struck 
      baseball hit at Coors Field as a function of the velocity off the bat. 
      Positive values of $\Delta x$ mean that the ``drier'' baseball
      falls short of the standard baseball. For ease of comparing to
      baseball units, the top axis is labeled in mph and the right axis in feet.}
  \end{center}
\end{figure}

A batted baseball's velocity off the bat depends on the initial speed of
the pitch, the speed of the bat, the rotation of the pitch, the
moment of inertia of the ball, and the impact parameter.\cite{nathan} 
Here, for simplicity, we consider an
optimally struck curveball, which leaves the bat with a speed of
about $40$ m/s at an angle of 24.3$^\circ$ from the
horizontal.\cite{shs} In our calculations, we assume a range of
initial speeds $35$-$45$ m/s off the bat at this angle.  Fig. \ref{rangevar}
shows the difference $\Delta x = x^{\rm s} - x^{\rm d}$, where $x^{\rm s}$ is the
distance the standard ball travels, and $x^{\rm d}$ is the distance the
dry ball travels.

In all cases, the standard ball actually travels slightly farther,
but only by about 2 feet.  Similar to the effect on lift, the 
acceleration due to drag is proportional to the ball's area, but 
inversely proportional to its mass. Since the fractional change in 
mass is greater than that in area, the mass effect wins. We apply the same 
ideas as in Sec. \ref{pitches} to find how the acceleration 
due to drag is affected by a change in RH. We define 
$\Delta a_{\rm D} = a_{\rm D}^{\rm s}-a_{\rm D}^{\rm d}$, where once again 
s and d refer to standard and dry baseballs. The acceleration due to drag is 
\begin{equation}\label{adrag}
  a_{\rm D} = -\frac{1}{2}\frac{\rho A}{m}C_{\rm D} v^2.
\end{equation}
A positive value of $\Delta a_{\rm D}$ implies the drier ball 
experiences more drag. To find the fractional change for a fixed initial speed 
we write
\begin{equation}
  \frac{\Delta a_{\rm D}}{a_{\rm D}^{\rm s}} = \frac{\Delta C_{\rm D}}{C_{\rm D}^{\rm s}}
  + \frac{\Delta A}{A^{\rm s}} - \frac{\Delta m}{m^{\rm s}}.
\end{equation}
Notice that if $\Delta a_{\rm D}/a_{\rm D}^{\rm s}<0$, the drier baseball will 
experience a fractionally larger acceleration due to drag. Using the data from the 
experiment, and assuming for the moment that $\Delta C_{\rm D}=0$, 
we find that the value of $\Delta a_{\rm D}/a_{\rm D}^{\rm s}\approx -1.12$\% 
for balls initially leaving the bat. Therefore, based solely on the change in the ball, 
one would expect humidified balls to travel slightly farther than dry balls, for the 
same initial conditions.

This analysis is complicated by the fact that the velocity, and thus $C_{\rm D}$, both 
change during the ball's flight. The value of 
$\Delta C_{\rm D}/C_{\rm D}^{\rm s}\ne0$ over the whole range of launch velocities for 
the drag profiles of SHS and Frohlich. For SHS, $\Delta C_{\rm D}/C_{\rm D}^{\rm s}$ 
monotonically decreases from zero to small negative values while for Frohlich's model 
we find that it increases from small negative values to small positive values. 
These can drastically change the amount of acceleration due to drag 
experienced by the ball during its flight. In order to understand the complete 
effect of the variation of the drag coefficient a full trajectory calculation 
is needed, and this is what is presented in Fig. \ref{rangevar}. We can see that 
the monotonically decreasing value of $\Delta C_{\rm D}/C_{\rm D}^{\rm s}$ in the 
SHS drag profile causes the slope in the range variation to increase with greater launch 
velocity, thus deviating from the values of Nathan and a smooth sphere. In the curve 
of Frohlich, we see the slope in range variation diminish and this is due to the change 
in sign of $\Delta C_{\rm D}/C_{\rm D}^{\rm s}$ that occurs around $40$\;m/s.

Yet, we know that RH has other effects on the batted ball beyond aerodynamics. 
For example, Kagan\cite{kag2} found that a 20\% increase 
in RH would reduce the ball's coefficient of restitution, resulting in 
about a 6 foot reduction on the distance of a batted ball. This change 
represents the net effect of two competing tendencies. The standard ball comes off the 
bat slower than the dry one, and so would not go as far. However, the slower ball also 
experiences less deceleration due to drag, and might be expected to go farther. Only 
the full trajectory calculation can tell which effect dominates, as pointed out by Kagan. 
It turns out that it is the velocity off the bat that dominates and therefore the 
standard ball should travel less far. One can naively expect the total change in range 
to be the sum of the collisional and aerodynamic effects, which yields about a $4$ 
foot overall reduction in batted range. This is the subject of the following section. 
In the following subsection we will see that this is so.

\subsection{Net Effect of RH on Batted Baseballs}

Nathan and Cross\cite{cnr} have worked out the way in which the bat-ball collision 
depends on the parameters $e$, $d$ and $m$ - the coefficient of restitution, the diameter 
of the ball and the mass of the ball. Using the measurements of Kagan\cite{kag2} and 
the current work, we can extract the changes in launch parameters such as angle, speed and 
spin-rate of the ball. In order to do so, we have to clearly define the assumptions made 
before the collision. We assume the ball arrives from the pitcher in the exact same location 
with the same velocity and spin-rate for a dry and standard ball and that the hitter takes 
the same swing with the bat to the ball. Therefore, the only parameters that change in the 
collision due to RH changes are $e$, $d$ and $m$ off the ball.

\begin{figure}[h]
  \begin{center}
    \includegraphics{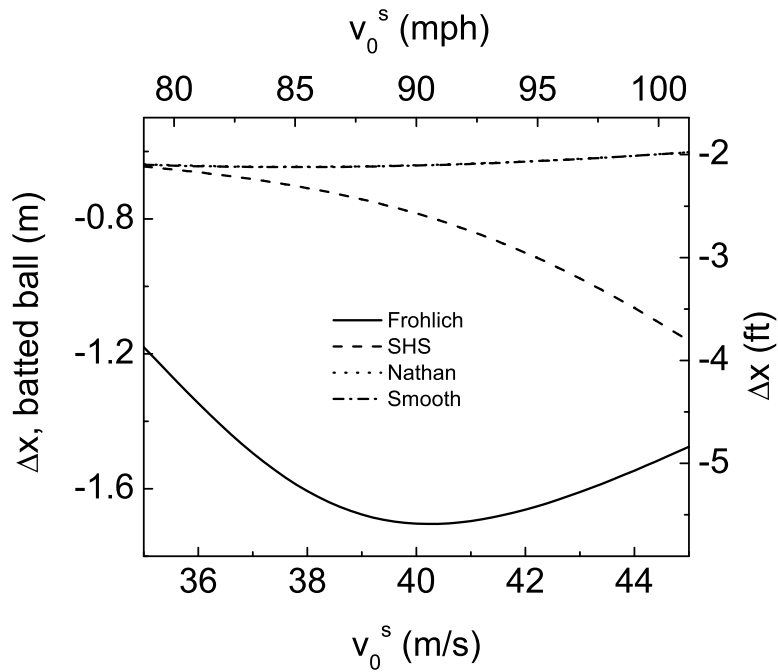}
    \caption{\label{totalvar} Relative variation in the range of a well-struck 
      baseball hit at Coors Field as a function of the velocity off the bat including 
      the effects of an increased coefficient of restitution. 
      Negative values of $\Delta x$ mean that the ``drier'' baseball
      flies further than the standard baseball. For ease of comparing to
      baseball units, the top axis is labeled in mph and the right axis in feet.}
  \end{center}
\end{figure}

We present in Fig. \ref{totalvar} the sum total of the effects of RH on the range of 
a batted ball including the effects of the collision. The range variation 
is not quite the 4 feet expected from adding the collisional and aerodynamic effects 
independently, presumably due to the non-linear interaction of the variables involved. 
The results for the smooth ball and the one of Nathan are fairly constant, although they 
start to deviate slightly as the launch velocity is increased. 
The small diminution in the range variation is due to the $v^2$ of the deceleration due 
to drag. Balls leaving with a larger velocity off the bat experience more deceleration. The 
increase in range variation of the SHS profile is due to the large change in $C_{\rm D}$ 
experienced by the ball during its trajectory; though only the left side (smaller 
$\mathcal{R}$ values) of the crisis is sampled. The drag profile of Frohlich samples both 
the drop and rise of $C_{\rm D}$ and therefore there is a maximum in the range variation. 
This occurs near $40$\;m/s.

Again, a more detailed study of the drag profile for spinning baseballs is needed to 
fully understand the effects of RH variation on the game of baseball. At sea-level, these 
effects will be more pronounced, just as in the case of pitching. At Coors Field, it seems 
these RH variations are a small effect on the range of a batted baseball, $\sim 3$\;feet, 
an average of the values for a standard ball launched at $40$\;m/s. This value is consistent 
with what is observed (Table \ref{data}). A more detailed study that includes the 
prevailing northeast winds\cite{chambers}, a breakdown of flyballs to left, left-center, 
center, right-center and right field, and a 3 dimensional model are needed to truly nail 
down range variations. The fly ball data listed is an average over all field directions 
and balls hit to left field will be more affected due to the wind patterns in Coors Field 
than those hit to right field.

\section{Conclusions}\label{con}

We have shown that storing baseballs in humidity controlled
environments can slightly increase their size, mass, and density.
This change, in turn, has small consequences for the aerodynamics of
a ball in flight. Based solely on aerodynamic considerations, humidified balls will 
break slightly more, by about $0.2$ in, while batted balls may travel further, 
on the order of a couple of feet. Both effects appear at first counter-intuitive, but
follow from the aerodynamics of spheres in air. Both effects are also 
common in experience: a baseball and whiffle ball are the same size, 
but the denser baseball curves far less and can be batted much further. 
Moreover, the effect on the batted ball is largely negated by the decreased
coefficient of restitution of the humidified ball, causing an 
overall reduction of a few feet in the distance a ball travels. In addition to 
collisions and aerodynamics, other effects may contribute to the humidor's influence 
on the game of baseball. An intriguing possibility is that the humidified balls are 
easier to grip, allowing pitchers to put a greater spin on a humid ball than on a
dry one.

\begin{acknowledgments}
This work was supported by the NSF.  We acknowledge Diana and Bob Meyer 
for providing the initial sample of baseballs, and  David
Alchenberger and Debbie Jin for technical assistance.
\end{acknowledgments}

\bibliography{refs}

\end{document}